# Automated Instrumentation for High-Temperature Seebeck Coefficient Measurements

Ashutosh Patel[1] and Sudhir K. Pandey[1]

[1]School of Engineering, Indian Institute of Technology Mandi, Kamand 175005, Himachal Pradesh, India

Corresponding author: Ashutosh Patel, E-mail: ashutosh_patel@students.iitmandi.ac.in

**Abstract**

In this work, we report the fabrication of fully automated experimental setup for high temperature Seebeck coefficient ($\alpha$) measurement. The K-type thermocouples are used to measure the average temperature of the sample and Seebeck voltage across it. The temperature dependence of the Seebeck coefficients of the thermocouple and its negative leg is taken care by using the integration method. Steady state based differential technique is used for $\alpha$ measurement. Use of limited component and thin heater simplify the sample holder design and minimize the heat loss. The power supplied to the heater decides the temperature difference across the sample and the measurement is carried out by achieving the steady state. The LabVIEW based program is built to automate the whole measurement process. The complete setup is fabricated by using commonly available materials in the market. This instrument is standardized for materials with a wide range of $\alpha$ and for the wide range of $\Delta T$ across the specimen. For this purpose, high temperature measurement of $\alpha$ of iron, constantan, bismuth, and $Bi_{0.36}Sb_{1.45}Te_3$ samples are carried out and data obtained from these samples are found to be in good agreement with the reported data.

# INTRODUCTION

In the present century, it is a major challenge to fulfill the demand of electricity for everyone. Currently maximum part of electricity is generated by using various types of heat engines like subcritical coal fired power station (maximum efficiency = 42 % [1]), supercritical coal fired power station (maximum efficiency = 49 %), etc. So more than half of suppling energy is released directly to the environment as waste heat. This waste heat can be converted into electricity by using themoelectric generators. The conversion efficiency of themoelectric materials depends on the Figure-of-merit (ZT). The larger value of ZT, higher will be the efficiency of material.[2] ZT of any themoelectric material is calculated by using the following formula,

$$ZT = \alpha^2 T/\rho\kappa \qquad (1)$$

Where $\alpha$, $\rho$, $\kappa$, and T are Seebeck coefficient, electrical resistivity, thermal conductivity, and absolute temperature, respectively. From above equation, it is clear that the ZT is proportional to square of $\alpha$ and hence it plays an important role in calculating the value of ZT. The Seebeck coefficient depends on transport properties of charge carriers and thus affected by impurity, defects, and phase transformation in materials.[3] So, we need an instrument which should be capable of measuring the $\alpha$ in a wide temperature range with fairly good accuracy at low cost and capable to characterize a wide variety of samples with various dimensions.

The Seebeck coefficient is determined in two ways: integral and differential methods. In integral method, one end of the sample is kept at a fixed temperature ($T_1$) and temperature of the other end ($T_2$) is tuned to the desired range. The Seebeck voltage generated in the sample is recorded as a function of temperature $T_2$. The Seebeck coefficient of the sample relative to the connecting wire at any temperature can be obtained from the slope of the Seebeck voltage versus temperature curve at that temperature. A large Seebeck voltage is generated due to the large temperature difference across the sample ($\Delta T$). This large Seebeck voltage minimizes the error due to the presence of small spurious voltage generated in the measurement circuit.[4] The requirement of additional cooling system to keep one end of sample at fixed temperature increases the complexity of the instrument along with the cost. Integration method is not applicable for nondegenerate semiconductors and insulators.[5, 6]

In differential method, Seebeck coefficient is calculated by the given equation,

$$\alpha_s = -\frac{\Delta V}{\Delta T} + \alpha_w \qquad (2)$$

Where $\alpha_s$, $\Delta V$, and $\alpha_w$ are absolute Seebeck coefficient of the sample, measured Seebeck voltage, and Seebeck coefficient of connecting wire, respectively. This is the conventional method of Seebeck coefficient measurement. There is no need of any additional cooling system in this method. Differential method is suitable for any type of materials. Due to the reason described above, the differential method is used in the most of the instruments for $\alpha_s$ measurement.[7-11] Low Seebeck coefficient materials (copper, niobium, and platinum) are used as connecting wire to measure Seebeck voltage and thermocouples are used to measure temperatures. Inherent accuracy of thermocouple may also lead to inaccuracy in temperature measurement.[12] Inaccurate measurement of temperatures may also cause inaccuracy in $\Delta T$ measurement, which can change the $\alpha_s$ largely in the case of low $\Delta T$. To measure the temperature and Seebeck voltage, connecting wire and thermocouple should be fixed at the exactly same point of the sample, which is a major difficulty.[6]

To overcome these limitations, *Boor et al.* suggested a different equation given below, which can be derived by using the conventional equation.

$$\alpha_s = -\frac{U_{neg}}{U_{pos} - U_{neg}} \alpha_{TC} + \alpha_{neg} \qquad (3)$$

Where $U_{pos}$ and $U_{neg}$ are Seebeck voltages measured by using positive legs and negative legs of thermocouple wires, respectively. $\alpha_{TC}$, and $\alpha_{neg}$ are Seebeck coefficient of thermocouple and its negative leg, respectively. In thisequation, Seebeck voltage is measured using thermocouple legs only and no additional wires are required. This resolve the difficulty in the measurement of temperature and Seebeck voltage at exactly the same point of sample. The use of thermocouple as connecting wire also simplified the design of the instrument. Employing Eqn. 3 over Eqn. 2 is highly advantageous for several reasons. Firstly, spurious thermal offset voltages from the system are cancelled. Secondly, the equation requires no direct temperature measurements, which tend to be less accurate than voltage measurements.[13] Temperature measurements are required only to find the value of $\overline{T}$ (mean temperature), $\alpha_{TC}$, and $\alpha_{neg}$, where accuracy is less important. Extra effort is required to find the value of $\alpha_{TC}$, and $\alpha_{neg}$. Eqn. 3 is used in very few papers for $\alpha_s$ measurement.[13-16] Out of them *Kolb et al.*[15] and *Boor et al.*[13, 16] found the value of $\alpha_{TC}$ and $\alpha_{neg}$ using the equations, given below

$$\alpha_{TC}(T_C, T_H) \approx \alpha_{TC}(\overline{T}) \qquad (4)$$

$$\alpha_{neg}(T_C, T_H) \approx \alpha_{neg}(\overline{T}) \qquad (5)$$

Above approximation is valid in two cases (i) $\Delta T$ should be small[13,15,16] and (ii) Seebeck coefficient of the positive and negative legs should have linear temperature dependence. For wide temperature range, case (ii) is difficult to satisfy so, Eqn. 3 is limited to the small $\Delta T$ with above approximation. Maintaining a constant $\Delta T$ throughout the experiment requires an additional heater at cold side and temperature controller. This also makes the measurement process complex and increases the cost of setup.

In the present work, we have addressed above issues and developed a low cost fully automated instrument to measure $\alpha_s$. The Seebeck coefficient of thermocouple and its negative leg has been calculated by integrating the temperature dependent values of $\alpha_{TC}$ and $\alpha_{neg}$. A single thin heater is used to heat the sample and $\Delta T$ across the sample is generated due to its thermal conductivity. This heater provides high temperature at low power supply compared to bulk heater. Italso simplifies the sample holder design due to its small size. The sample holder is lightweight and small in size, in which limited components are used. Each component of the sample holder is fabricated separately, which provides us liberty to replace any part if it gets damaged. Its simple design makes loading and unloading of the sample easier and is capable of holding samples of various dimensions. The LabVIEW based program is built to automize the measurement process. Iron, constantan, bismuth and $Bi_{0.36}Sb_{1.45}Te_3$ samples are used to validate the instrument. The data collected on these samples are found to be in good agreement with the reported data.

## PRINCIPLE OF MEASUREMENT

Theoretically, the Seebeck voltage across a sample ($V_s$) can be written as,

$$V_s(T_C, T_H) = -\int_{T_C}^{T_H} \alpha_s(T) dT \tag{6}$$

Where $T_H$, $T_C$ and $\alpha_s(T)$ are hot side temperature, cold side temperature, and Seebeck coefficient as a function of temperature.

Connecting wires are required to measure Seebeck voltage across the sample. The temperature difference is also generated across both connecting wires, which adds its own Seebeck voltage in the measured voltage. The free ends of both connecting wires are at temperature $T_R$.

The expression for measured voltage ($V_m$) in-terms of sample voltage ($V_s$), cold side wire voltage ($V_{wc}$), and hot side wire voltage ($V_{wh}$) can be written as,

$$V_m(T_C, T_H) = V_{wc} + V_s + V_{wh} \tag{7}$$

By using Eqn. 6, we can write the expression for $V_{wc}$ and $V_{wh}$ as,

$$V_{wc}(T_R, T_C) = -\int_{T_R}^{T_C} \alpha_w(T) dT \tag{8}$$

$$V_{wh}(T_H, T_R) = -\int_{T_H}^{T_R} \alpha_w(T) dT \tag{9}$$

Now, by putting the values of $V_{wc}$ and $V_{wh}$ from Eqns. 8 and 9, respectively in Eqn. 7, we get

$$V_m(T_C, T_H) = V_s + V_{wc} + V_{wh}$$

$$= V_s - \int_{T_R}^{T_C} \alpha_w(T)dT - \int_{T_H}^{T_R} \alpha_w(T)dT$$

$$= V_s - \int_{T_R}^{T_C} \alpha_w(T)dT + \int_{T_R}^{T_H} \alpha_w(T)dT$$

$$= V_s - \int_{T_R}^{T_C} \alpha_w(T)dT + \int_{T_R}^{T_C} \alpha_w(T)dT + \int_{T_C}^{T_H} \alpha_w(T)dT$$

$$= V_s + \int_{T_C}^{T_H} \alpha_w(T)dT \tag{10}$$

If $V_w$ is net Seebeck voltage due to both connecting wires, then from Eqn. 10, it can be written as,

$$V_w(T_C, T_H) = -\int_{T_C}^{T_H} \alpha_w(T)dT$$

So, the effective Seebeck coefficient of connecting wire can be written as

$$\alpha_w(T_C, T_H) = -\frac{1}{T_H - T_C}\int_{T_C}^{T_H} \alpha_w(T)dT \tag{11}$$

Using the above equation, we can calculate $\alpha_w$ accurately.

Thermocouple is having the same boundary condition like connecting wire. So, the equation derived above can be written for thermocouple also.

$$\alpha_{TC}(T_C, T_H) = -\frac{1}{T_H - T_C}\int_{T_C}^{T_H} \alpha_{TC}(T)dT \tag{12}$$

For conventional method, only $\alpha_w$ is required, while for equation suggested by *Boor et al.*, $\alpha_w$ as well as $\alpha_{TC}$ are required. $\alpha_w$ represents the Seebeck coefficient of connecting wire, if negative leg of thermocouples used, it will be written as $\alpha_{neg}$. In this work, from now onward Seebeck coefficient of connecting wire is indicated as $\alpha_{neg}$. It is clear from the above discussion that the method proposed in the present work is expected to be better than that of *Boor et al.* for the general type of thermocouples and for a large value of $\Delta T$.

## MEASUREMENT SETUP

A schematic view of measurement setup is shown in Fig. 1. The digital multimeter with the multichannel scanner card is used to measure various signals. Sourcemeter is used to supply power to the heater. GPIB ports of the digital multimeter and Sourcemeter connected by using inline IEEE-488 GPIB bus interface cable. GPIB-USB converter is used to connect GPIB with a computer. Digital multimeter measures $U_{pos}$, $U_{neg}$, $T_H$, $T_C$, and connector's temperature ($T_{ref}$). $T_{ref}$ is considered as cold junction compensation temperature for thermocouples and measured by using PT-100 RTD. Shielded cable is used to avoid electrical noise due to inductive coupling in signal transmission from connector to scanner card.

The detailed overview of sample holder assembly is shown in Fig. 2, where different components are represented by numbers. The sample (1) is sandwiched in between two copper blocks (2) of 10mm×10mm cross section and 2 mm thickness. The two K-type polytetrafluoroethylene coated thermocouples (3) of 36 swg are embedded in the copper blocks. In order to make a good thermal and electrical contact between copper block and thermocouple GaSn liquid metal is used. Fine thermocouple wire minimizes the heat flow through thermocouple, which helps to measure more accurate temperature. Thin heater (4) is used to heat the sample and it is made by winding 40 swg kanthal wire over the mica sheet and wrapped by using another mica sheet then by copper sheet. The cross section of this heater is 10mm×10mm and thickness is 1 mm. Thick insulator block (5) is placed in between the heater and brass plate (6) to minimize heat loss. The cross section and thickness of insulator block are 10mm×10mm and 25 mm, respectively. Another insulator block (7) is used in between cold side copper block and fully threaded stainless steel rod (8) to insulate sample electrically from surrounding. This stainless steel rod is supported by a second brass plate (9) and used to apply pressure on the sample. This applied pressure, ensures a good surface contact between sample and copper blocks. Both brass plates are of same dimension 40mm×15mm cross section and 8 mm thickness. Both brass plates are fixed over another stainless steel rod (10) by using the screws ((11) & (12)). Both stainless steel rods are about 6 mm diameter and 100 mm length. This stainless steel rod is fixed over stainless steel flange (13). On this stainless steel flange, hermetically sealed electrical connector (14) is fixed to make electrical connections. This connector is also acting as cold junction for thermocouples. PT-100 RTD (15) is used to measure the temperature of cold junction. This measured temperature is taken as cold junction compensation for thermocouples. Vacuum chamber (16) is made by using seamless stainless steel pipe of 10 cm diameter and 30 cm in height. KF25 port (17) is provided over vacuum chamber. This port is used to connect the vacuum chamber with vacuum pump. Rotary vane pump is used to create a vacuum inside the chamber upto a level of 0.018 mbar.

To control the whole measurement process, a program is built on LabVIEW programming language. Before starting the program, the user needs to fill sample details and other control parameters (step increment in the power supply, number of data average, and $T_H$ limit). Instead of controlling temperature, we controlled power supply and based on supplying power a temperature is achieved. Once steady state is achieved digital multimeter starts acquiring data. The random error generated in measurement is minimized by taking an average of multiple data. Fast data acquisition takes place with the help of the GPIB based interface system.

The method of finding the value of $\alpha_{TC}$ and $\alpha_{neg}$ described in Eqn. 11 and 12 are implemented by making a program in LabVIEW. This program requires polynomial coefficients of temperature dependent Seebeck coefficient function, $T_C$, and $T_H$. In order to find the polynomial coefficient, we have used data reported in Refs. 16 and 17 for alumel and K-type thermocouple, respectively. The data fitted with polynomial equations of degree 19. The polynomial coefficients, thus obtained are used to estimate the value of temperature dependent $\alpha_{TC}$ and $\alpha_{neg}$. The measured value of $U_{pos}$ and $U_{neg}$ along with estimated $\alpha_{TC}$ and $\alpha_{neg}$ are used in Eqn. 3 which gives the value of $\alpha_s$ at a given temperature. This $\alpha_s$ is plotted online with temperature and all raw data along with $\alpha_s$ is exported to .xls file. After completion of this measurement loop, sourcemeter increases the power supply to heater by the value defined in control parameter. This measurement loop continues until $T_H$ reaches a set value defined in control parameter. Once $T_H$ reaches set value, sourcemeter stops power supply to the heater.

## RESULTS AND DISCUSSIONS

The instrument is validated by measuring the temperature dependent Seebeck coefficient of constantan, iron, bismuth, and $Bi_{0.36}Sb_{1.45}Te_3$ samples. Measurements of various samples are carried out to demonstrate the flexibility of the instrument. Atthis point, we would like to re-emphasize the fact that, in this instrument $\Delta T$ is not controlled. It is generated due to the heat flow through the sample, sample dimension, its thermal conductivity ($\kappa$), and thermal contact resistance between sample, heat loss through insulator, and copper block, but still this $\Delta T$ range can be tuned by tuning the sample dimensions or insulator dimension or by selecting a lowest thermal conductive material as insulator block. In order to check the suitability of our integration method used to calculate $\alpha_{TC}$ and $\alpha_{neg}$ with a different $\Delta T$ range, we performed measurement with three different temperature ranges on iron and constantan samples. For this, samples of different dimensions are prepared by turning wires of iron and constantan extracted from J-type thermocouple into multilayer flat spiral shape of different thickness. These samples are symbolically indicated by Fe1, Fe2, Fe3 and C1, C2, C3 for iron and constantan, respectively. The more details of these samples are given in table 1. In this setup, 5 watts of power supply is sufficient to get the hot side temperature of 650 K.

The variation in $\Delta T$ with $\overline{T}$ for Fe1, Fe2, and Fe3 samples are shown in Fig. 3. At 312 K, $\Delta T$ is nearly 0.25 K, 1.6 K, and 9 K for Fe1, Fe2, and Fe3, respectively. The value of $\Delta T$ are increasing with temperature and reachto 16 K at $\overline{T}$=625 K for Fe1, 56 K at $\overline{T}$=585 K for Fe2, and 160 K at $\overline{T}$=540 K for Fe3 samples. The rate of change of $\Delta T$ is very high for Fe3 compared to Fe1 and Fe2 samples. The vlues of $\alpha$ for Fe1, Fe2, and Fe3 samples with respect to $\overline{T}$ are shown in Fig. 4. The values of $\alpha$ for all the three samples are decreasing almost linearly throughout the temperature range, which is as per the reported data[17]. Below 400 K, small difference (maximum ~0.5 $\mu V/K$) is observed among the measured values of $\alpha$ for Fe1, Fe2 and Fe3 samples. At 400 K, the values of $\alpha$ for Fe1, Fe2 and Fe3 samples match closely with each other. At this temperature, $\Delta T$ for Fe1, Fe2, and Fe3 are 2.5 K, 17 K, and 54 K, respectively. After this temperature, again the difference in the values of $\alpha$ for all the three samples increases. Upto 500 K, maximum deviation of 1 $\mu V/K$ is observed among the measured values of $\alpha$ for Fe1, Fe2 and Fe3 samples. Due to large $\Delta T$ at high temperature, the value of $\alpha$ for Fe3 sample shows more deviation compared to Fe1 and Fe2 samples. At $\overline{T}$=525 K, the value of $\alpha$ for Fe3 sample shows difference of 0.9$\mu V/K$ and 1.8 $\mu V/K$ with respect to Fe1 and Fe2 samples, respectively. The values of $\alpha$ for Fe1 and Fe2 samples show a maximum difference of 1.2 $\mu V/K$ with each other till the end of the measurement. We also compared our data with the reported data available in Ref. 17. In the absence of clear information provided about the measurement method and $\Delta T$ range in Ref. 17, we compared our values of $\alpha$ for lowest $\Delta T$ range (Fe1 sample) with the reported data, which is shown in the Fig. 5. The values of $\alpha$ throughout the temperature range are almost parallel to the reported data and both decreases with a rate of ~0.045 $\mu V/K$. The deviation in our data is ~1 $\mu V/K$ at 315 K. This deviation increases upto ~1.9 $\mu V/K$ at 480 K and again decreases to ~1.2 $\mu V/K$ at the end of the measurement.

The values of $\Delta T$ with $\overline{T}$ for C1, C2, and C3 samples are shown in Fig. 6. At the start of measurement, $\Delta T$ is 0.35 K for C1, 2 K for C2, and 4 K for C3. Its values increase with temperature and reach to 26 K at $\overline{T}$=644 K for C1, 58 K at $\overline{T}$=630 K for C2, and 170 K at $\overline{T}$=525 K for C3. The rate of change of $\Delta T$ is very high for C3 compared to C1 and C2. The values of $\alpha$ for all the three samples with respect to $\overline{T}$ areshown in Fig. 7. The values of $\alpha$ for all the three samples increase almost linearly from the start of the measurement to ~435 K and after that slope changes which is as per the reported data[17,19]. Initially, the values of $ha$ for all the three samples match closely with each other. The difference in the values of $\alpha$ for all the three samples starts increasing with the increase in temperature. Below 525 K, the maximum difference observed in the value of $\alpha$ is 1.4 $\mu V/K$. This difference in $\alpha$ is appeared at $\overline{T}$=400 K, where $\Delta T$ is ~4.3 K for C1 and 18 K for C2, and 80 K for C3. At $\overline{T}$=525 K, the values of $\alpha$ for all the three samples match closely and measurement for C3 sample ends here. After this temperature, the difference in the value of $\alpha$ for C1 and C2 increases till 600 K to a value of 1.8 $\mu V/K$ and then decreases till the end of the measurement. We also compared our data with the reported data available in Refs. 17 and 19. We have collected data at different $\Delta T$ range. In Ref. 17, no clear information is provided about the measurement procedure. In Ref. 19, measurement was performed at constant $\Delta T$ (~3.5 K) throughout thetemperature range. So, the measured data at the lowest $\Delta T$ range (C1 sample) is compared with the reported data, which is shown in the Fig. 8. Our data show similar behavior compared to the both reported data. At 315 K, the deviation in our data is equal to ~2 $\mu V/K$ and ~3.5 $\mu V/K$ compared from Refs. 17 and 19, respectively. The deviation decreases with the increase in temperature. At 435 K, our data match closely with Ref. 19 while having a deviation of 1.7 $\mu V/K$ with the Ref. 17. After this temperature, the deviation increases till 480 K and at this temperature deviation is 1.4 $\mu V/K$ and 3 $\mu V/K$ compared from Ref. 17 and 19, respectively. Decreasing behavior in deviation is observed after this temperature. Our data match closely with both the reported data from 550 K.

It is important to note that, we used commercially available mica sheet in heater fabrication which is not advisable to go beyond 700 K. As there is some temperature gradient between the mica sheet and hot side temperature, we restrict hot side temperature to a maximum value of 650 K. Due to large $\Delta T$ for Fe3 and C3 samples, even $T_H \approx$625 K, $\overline{T}$ value appears below 550 K.

Now we take bismuth sample, as it has relatively higher Seebeck coefficient. This sample is taken from commercial ingot. The sample is cut into dimension of $9mm \times 7mm$ cross section and 12 mm thickness. The variation in $\Delta T$ along with $ineT$ is shown in the Fig. 9. At $\overline{T}$ =315 K, $\Delta T$ is ~2.5 K. $\Delta T$ increases almost linearly and reaches to 27 K at $\overline{T}$=440 K. The values of $\alpha$ of the bismuth sample with respect to $\overline{T}$ are shown in Fig. 10. At the start of the measurement, value of $\alpha$ is ~-64 $\mu V/K$. Value of $\alpha$ decreases almost linearly to ~-63 $\mu V/K$ at $\overline{T}$=415 K. After this temperature, slope changes and decreases to ~-61 $\mu V/K$ at $\overline{T}$=440 K. Due to the low melting temperature of bismuth, measurement was performed till $T_H$=455 K. We also compared our $\alpha$ with the reported data[20]. In Ref. 20, measurement was performed upto a temperature of 350 K only. We found similar behavior of our data, but the deviation is ~11 $\mu V/K$ with the reported data till 350 K. This deviation of our data may be due to the highly anisotropic nature of the electronic transport.[21] Since this sample is cheaper, we cannot expect that its purity will be upto a level of standard samples, and this may be an another aspect behind the deviation.

Finally, we consider $Bi_{0.36}Sb_{1.45}Te_3$ sample, which has a very high Seebeck coefficient (>200 $\mu V/K$) at room temperature. This sample is extracted from commercially available thermoelectric generator (TEC1-12706). The composition of the sample is obtained by performing Energy-dispersive X-ray spectroscopy. The sample is about $1.4mm \times 1.4mm$ cross section and 1.6 mm thickness. The variation in $\Delta T$ along with $\overline{T}$ is shown in the Fig. 11. At a temperature of $\overline{T}$=315 K, $\Delta T$ is ~11 K. $\Delta T$ increases almost linearly and reach to 105 K at $\overline{T}$=490 K. The values of $\alpha$ for $Bi_{0.36}Sb_{1.45}Te_3$ sample with respect to $\overline{T}$ are shown in Fig. 12. At the start of the measurement, the value of $\alpha$ is ~212 $\mu V/K$ and it increases to ~221 $\mu V/K$ at $\overline{T}$=375 K. After this temperature, the value of $\alpha$ decreases and reaches to ~160 $\mu V/K$ at $\overline{T}$=490 K. We also compared our data with the reported data[22]. In Ref. 22, sample was taken from commercially available $BiSbTe$ ingot. Our data show similar behavior compared to the reported data. We observed a deviation of ~10 $\mu V/K$ at $\overline{T}$=315 K. This deviation increases with the increase in $\Delta T$ and reaches to 20 $\mu V/K$ at $\overline{T}$=490 K where $\Delta T$ is ~105 K. This small difference in the magnitude of our data may be attributed to the presence of large $\Delta T$, as seen abovefor other samples.

## CONCLUSION

In this work, we have developed simple, low cost and fully automated experimental setup for Seebeck coefficient measurement. Average temperature of the sample and the Seebeck voltage across it were measured using K-type thermocouples. The temperature dependence of the Seebeck coefficients of the thermocouple and its negative leg used in Seebeck coefficient calculations was taken care by using the integration method. Thin heater, simple design and limited components make small size and lightweight sample holder. Temperature difference across the sample was decided based on the power supply to the heater. The LabVIEW based program makes the whole measurement process fully automated. Commonly available materials in the market were used in the fabrication of the complete setup. This setup is validated by using iron, constantan, bismuth and $Bi_{0.36}Sb_{1.45}Te_3$ samples with a wide range of Seebeck coefficient and wide range of temperature difference across it. The measured data were found in good agreementwith the reported data, which indicate that this instrument is capable to measure Seebeck coefficient with fairly good accuracy.

## ACKNOWLEDGEMENTS

The authors acknowledge R. S. Raghav and other workshop staffs for their support in the fabrication process of vacuum chamber and sample holder assembly.

# Tables

**Table 1.** The relevant properties of Fe1, Fe2, Fe3, C1, C2, and C3 samples.

**Table 1.**

| Symbol | Sample material | Shape | Thickness (t) order |
|--------|----------------|-------|---------------------|
| Fe1    | 36 swg iron wire extracted from J-type thermoocuple | Multilayer spiral coil | $t_{Fe1} < t_{Fe2} < t_{Fe3}$ |
| Fe2    |                |       |                     |
| Fe3    |                |       |                     |
| C1     | 36 swg constantan wire extracted from J-type thermoocuple | Multilayer spiral coil | $t_{C1} < t_{C2} < t_{C3}$ |
| C2     |                |       |                     |
| C3     |                |       |                     |

# Figure Captions

**Figure 1.** Schematic diagram of high temperature Seebeck coefficient measurement setup along with computer interface system.

**Figure 2.** Schematic diagram of sample holder assembly: (1) sample, (2) copper blocks, (3) K-type 36 swg polytetrafluoroethylene coated thermocouples, (4) thin heater, (5) thick insulator block, (6) brass plate, (7) another insulator block, (8) fully threaded stainless steel rod, (9) second brass plate, (10) another stainless steel rod, (11) & (12) screws, (13) stainless steel flange, (14) hermetically sealed electrical connector, (15) PT-100 RTD, (16) vacuum chamber, and (17) KF25 port.

**Figure 3.** Variation in temperature gradient with mean temperature of iron samples (Fe1, Fe2, and Fe3), where data corresponding to Fe1, Fe2, and Fe3 are shown by black circle, blue box, red triangle, respectively.

**Figure 4.** Variation in Seebeck coefficient with mean temperature of iron samples (Fe1, Fe2, and Fe3), where data corresponding to Fe1, Fe2, and Fe3 are shown by black circle, blue box, red triangle, respectively.

**Figure 5.** Variation of Seebeck coefficient versus mean temperature of iron sample (Fe1), where the data corresponding to Fe1 of present work and the data taken from handbook of temperature measurement[17] are shown by black circle and brown open star, respectively.

**Figure 6.** Variation in temperature gradient with mean temperature of constantan samples (C1, C2, and C3), where data corresponding to C1, C2, and C3 are shown by black circle, blue box, red triangle, respectively.

**Figure 7.** Variation in Seebeck coefficient with mean temperature of constantan samples (C1, C2, and C3), where data corresponding to C1, C2, and C3 are shown by black circle, blue box, red triangle, respectively.

**Figure 8.** Variation of Seebeck coefficient versus mean temperature of constantan sample (C1), where the data corresponding to C1 of present work, the data taken from handbook of temperature measurement[17], and the data reported by *Guan et. al.*[19] are shown by black circle, brown open star, and blue diamond, respectively.

**Figure 9.** Variation of temperature gradient with mean temperature of bismuth sample.

**Figure 10.** Variation of Seebeck coefficient of bismuth at different mean temperature.

**Figure 11.** Variation of temperature difference with mean temperature of $Bi_{0.36}Sb_{1.45}Te_3$ sample.

**Figure 12.** Variation of Seebeck coefficient of $Bi_{0.36}Sb_{1.45}Te_3$ at different mean temperature.

**Figure 1.**

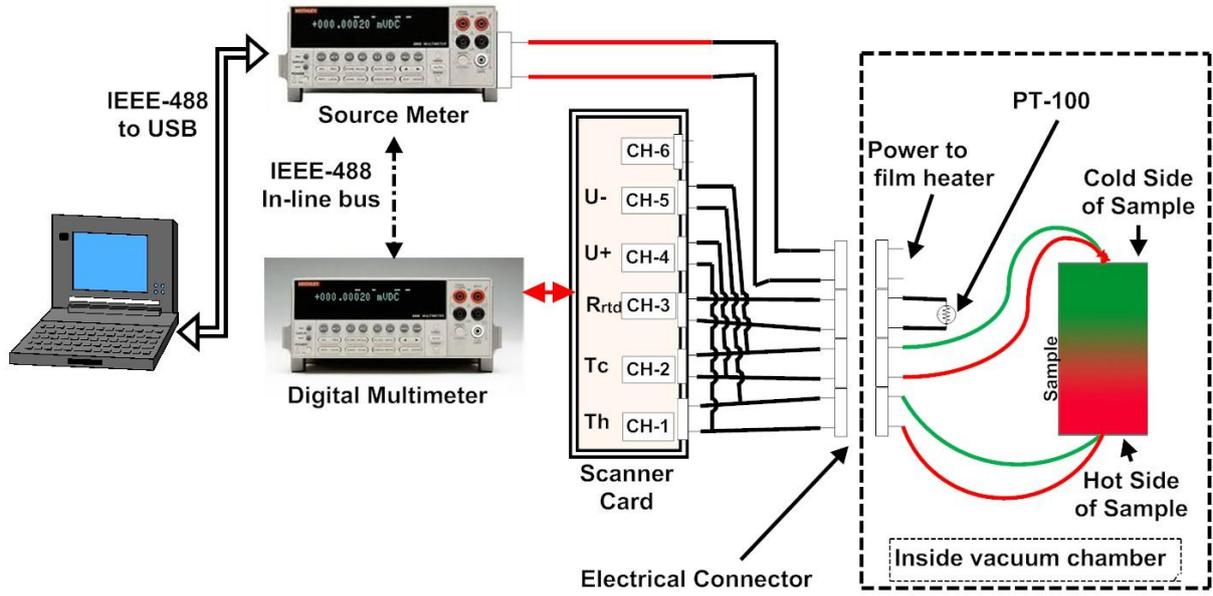

**Figure 2.**

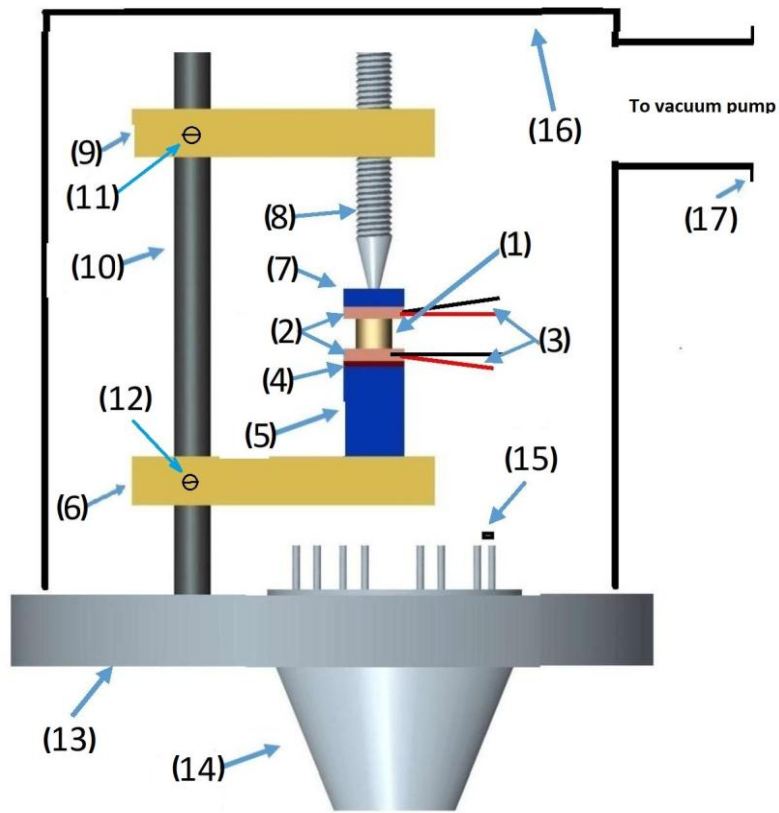

**Figure 3.**

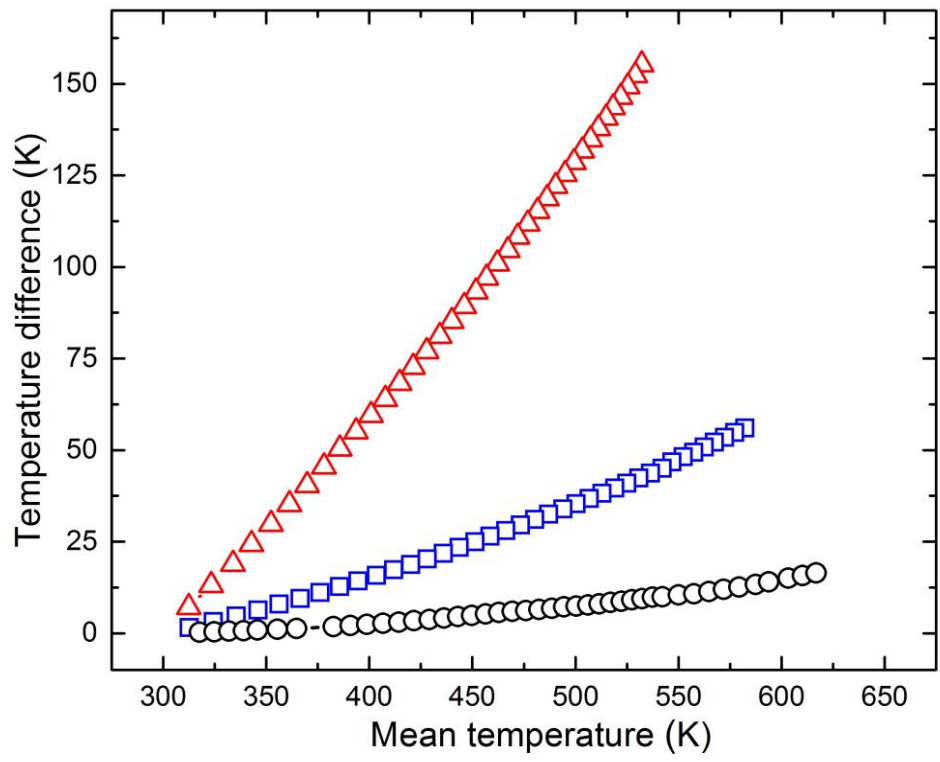

**Figure 4.**

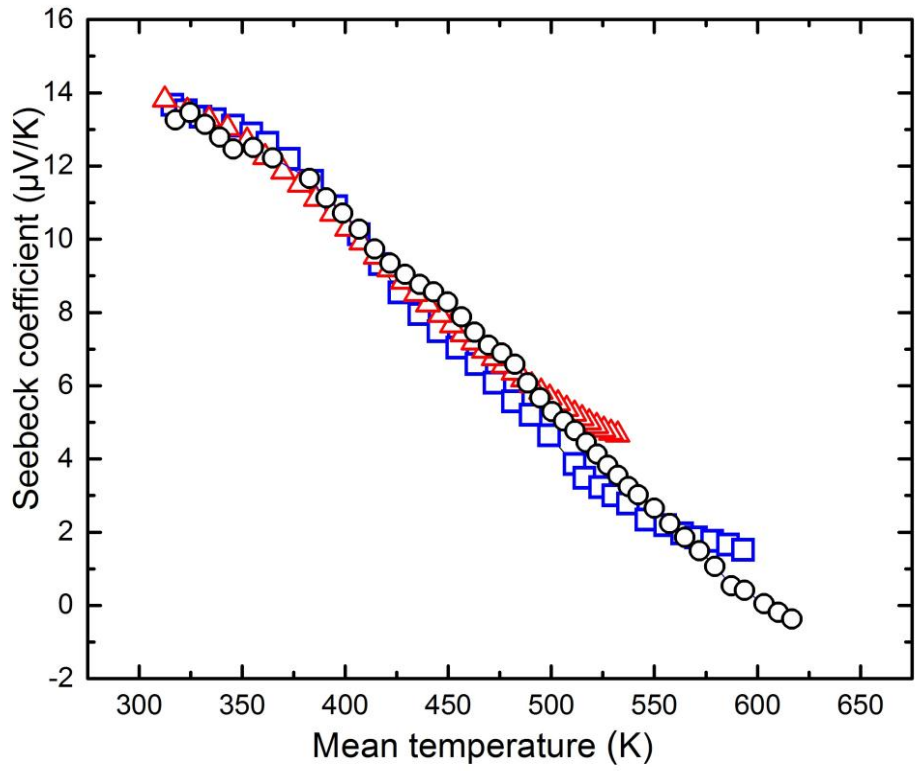

**Figure 5.**

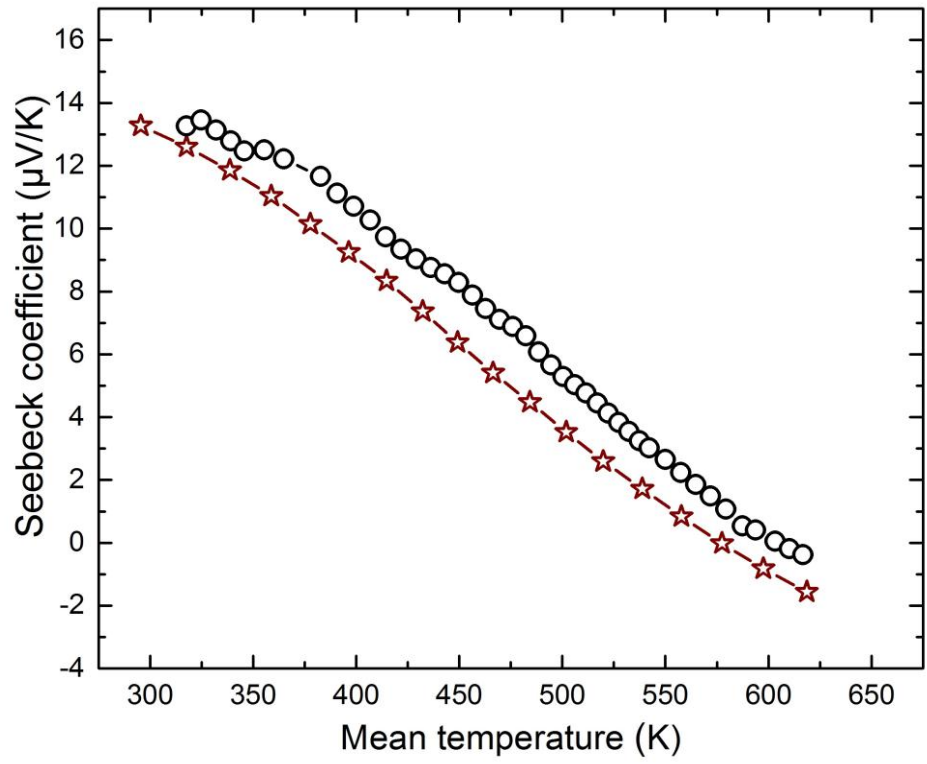

**Figure 6.**

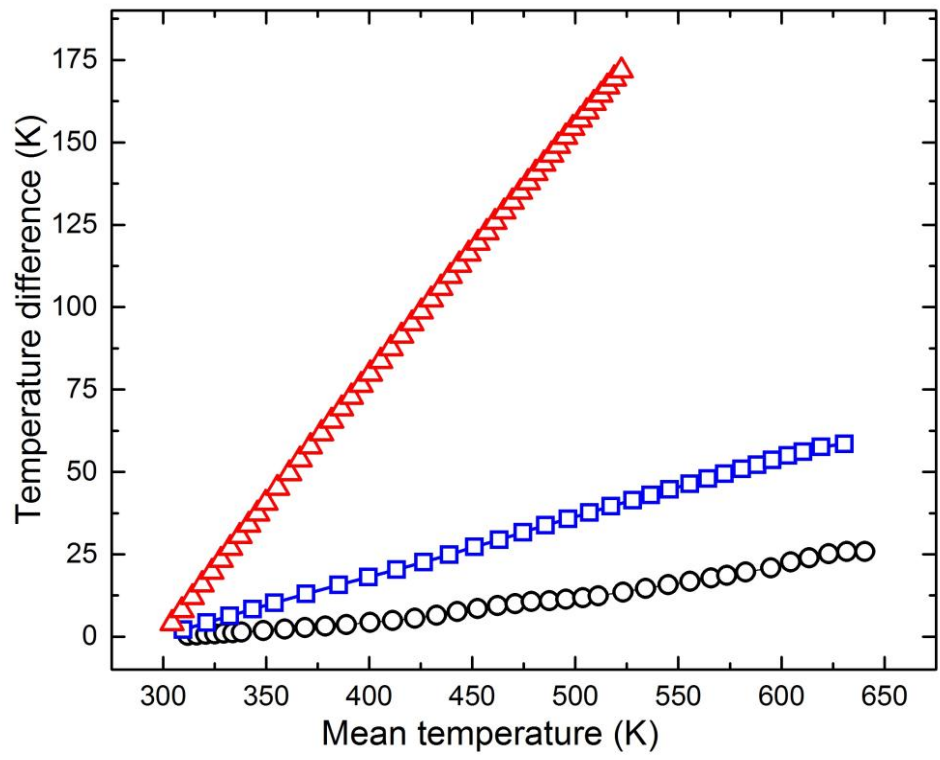

**Figure 7.**

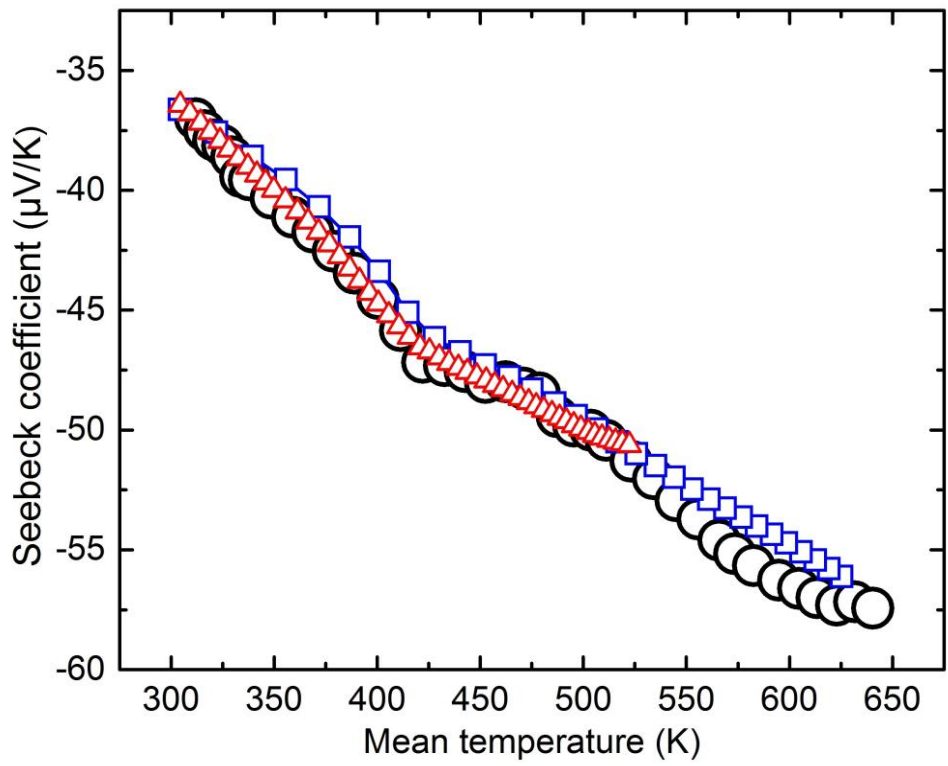

**Figure 8.**

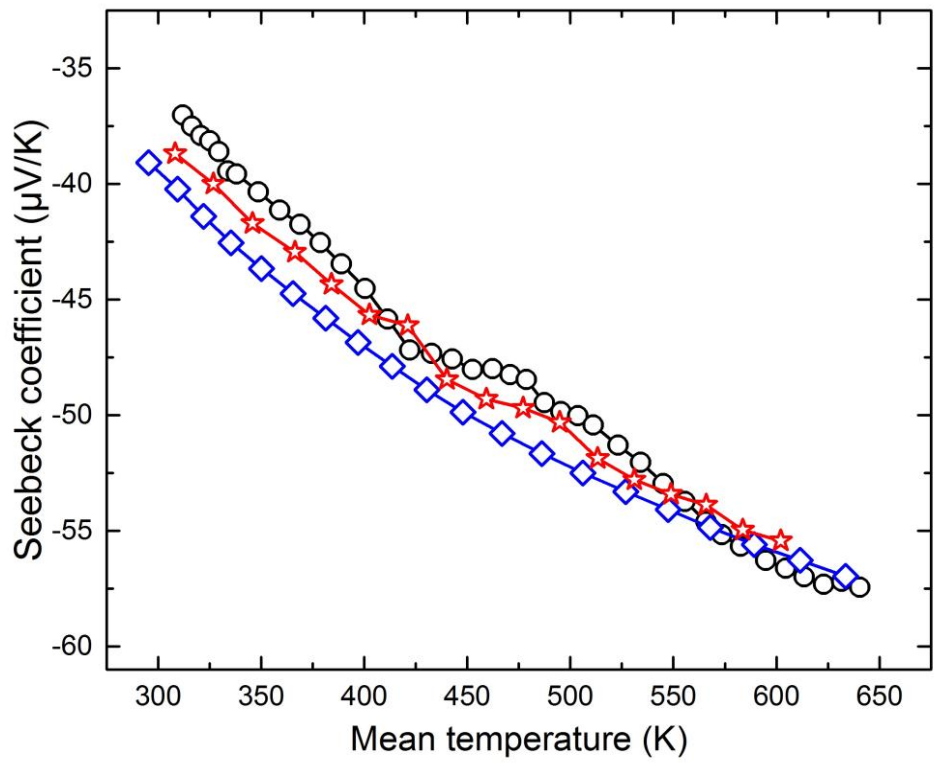

**Figure 9.**

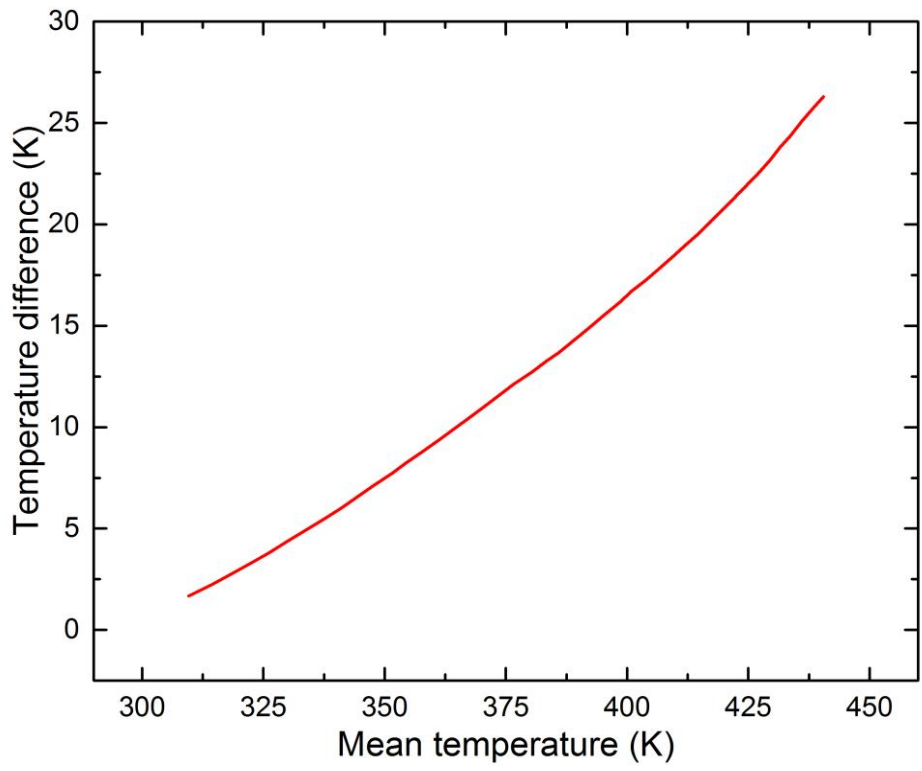

**Figure 10.**

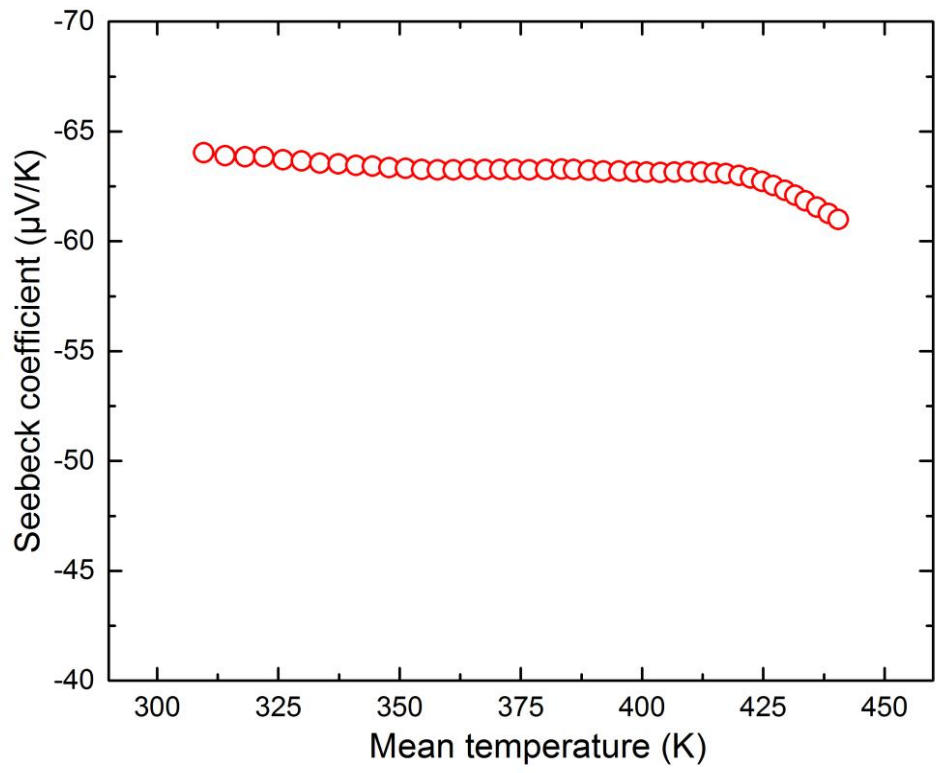

**Figure 11.**

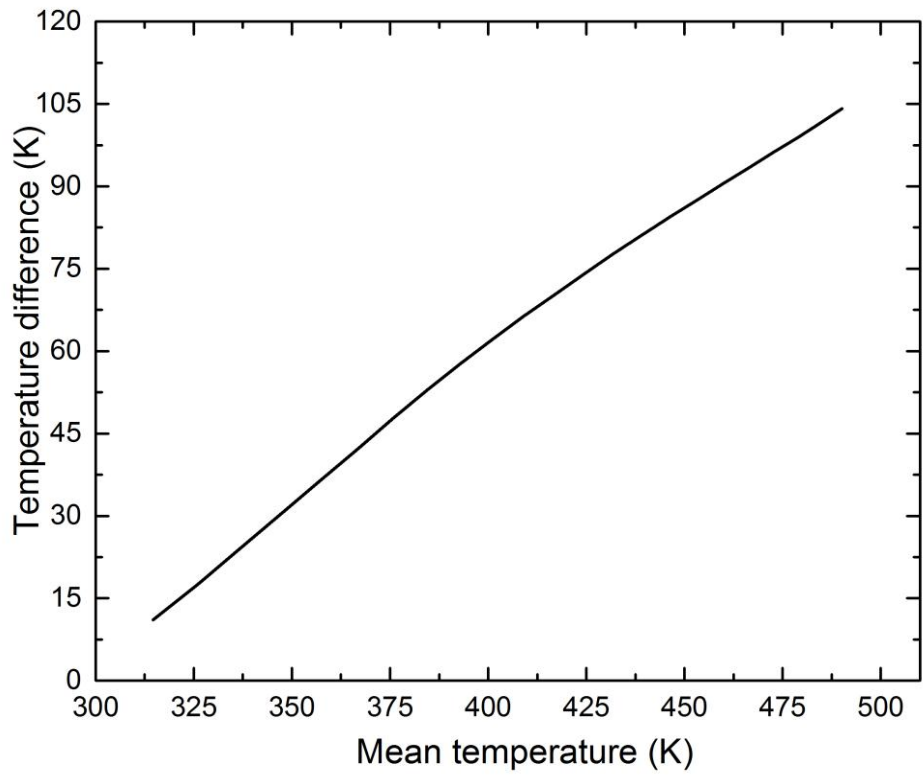

**Figure 12.**

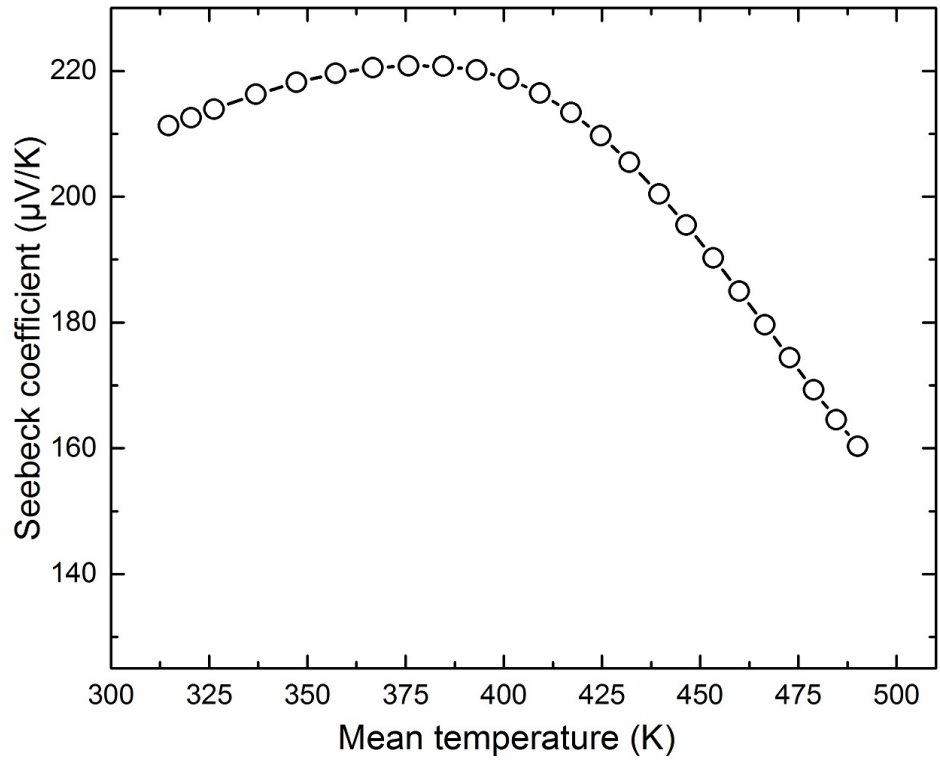